\documentstyle[12pt,epsf,epsfig]{article}
\makeatletter
\def\fmslash{\@ifnextchar[{\fmsl@sh}{\fmsl@sh[0mu]}}
\def\fmsl@sh[#1]#2{%
  \mathchoice
    {\@fmsl@sh\displaystyle{#1}{#2}}%
    {\@fmsl@sh\mboxstyle{#1}{#2}}%
    {\@fmsl@sh\scriptstyle{#1}{#2}}%
    {\@fmsl@sh\scriptscriptstyle{#1}{#2}}}
\def\@fmsl@sh#1#2#3{\m@th\ooalign{$\hfil#1\mkern#2/\hfil$\crcr$#1#3$}}
\makeatother

\begin{document}

\vspace{1.cm}
\begin{center}
\Large\bf Understanding the $D^+_{sJ}(2317)$ and $D^+_{sJ}(2460)$
with Sum Rules in HQET
\end{center}
\vspace{0.5cm}
\begin{center}
{ Yuan-Ben Dai$^1$, Chao-Shang Huang$^1$, Chun Liu$^1$, and
Shi-Lin Zhu$^2$}\\
\vspace{3mm} {\it $^1$Institute of Theoretical Physics,
 Chinese Academy of Sciences, P.O. Box 2735, Beijing 100080, China \\
  and \\
  $^2$Department of Physics, Peking University, Beijing 100871, China}
\end{center}

\vspace{1.5cm}
\begin{abstract}

In the framework of heavy quark effective theory we use QCD sum
rules to calculate the masses of the $\bar c s$ $(0^+, 1^+)$ and
$(1^+, 2^+)$ excited states. The results are consistent with that
the states $D_{sJ}(2317)$ and $D_{sJ}(2460)$ observed by BABAR and
CLEO are the $0^+$ and $1^+$ states in the $j_l={1\over 2}^+$
doublet.

\end{abstract}


\pagenumbering{arabic}

\section{Introduction}
\label{sec1}

Recently BaBar Collaboration announced a positive-parity narrow
state with a rather low mass $2317\pm 3$ MeV in the $D_s^+
(1969)\pi$ channel \cite{babar}, which was confirmed by CLEO
\cite{cleo} and BELLE \cite{belle2003} later. Because of its low
mass and decay angular distribution, its $J^P$ is believed to be
$0^+$. In the same experiment CLEO \cite{cleo} observed a state at
$2460$ MeV with the possible spin-parity $J^P=1^+$ in the
$D^\ast_s \pi$ channel. BaBar \cite{babar} also found a signal
near there. Since these two states lie below $D K$ and $D^\ast K$
threshold respectively, the potentially dominant s-wave decay
modes $D_{sJ}(2317) \to D_s K$ etc are kinematically forbidden.
Thus the radiative decays and isospin-violating strong decays
become favorable decay modes. The later decay goes in two steps
with the help of virtual $D_s \eta$ intermediate states,
$D_{sJ}(2317)\to D_s \eta \to D_s \pi^0$ where the second step
arises from the tiny $\eta$-$\pi^0$ isospin-violating mixing due
to $m_u \ne m_d$.

The experimental discovery of these two states has triggered
heated debate on their nature recently in literature. The key
point is to understand their low masses. The $D_{sJ}(2317)$ mass
is significantly lower than the values of $0^+$ mass in the range
of $2.4-2.6$ GeV calculated in quark models of \cite{qm}. The
model using the heavy quark mass expansion of the relativistic
Bethe-Salpeter equation in \cite{jin} predicted a lower value
$2.369$ GeV of $0^+$ mass which is still 50 MeV higher than the
experimental data. Bardeen, Eichten and Hill interpreted them as
the $\bar c s$ $(0^+, 1^+)$ spin doublet as the parity conjugate
states of the $(0^-, 1^-)$ doublet in the framework of chiral
symmetry\cite{bardeen}\footnote{The existence of parity doublets
has also been shown by combining chiral symmetry and heavy quark
symmetry in Bethe-Salpeter approach in \cite{huangcs}.}  (See also
Ref. \cite{ebert}). A quark-antiquark picture was also advocated
by Colangelo and Fazio \cite{colangelo}, Cahn and Jackson
\cite{cahn}, Godfrey \cite{godfrey}. Based on such a
"conventional" picture, the various decay modes were discussed in
Refs. \cite{bardeen,colangelo,godfrey}.

Apart from the quark-antiquark interpretation, $D_{sJ}(2317)$ was
suggested to be a four-quark state by Cheng and Hou \cite{cheng}
and Barnes, Close and Lipkin \cite{barnes}. Szczepaniak even
argued that it could be a strong $D \pi $ atom \cite{szc}! But we
think it would be very exceptional for a molecule or atom to have
a binding energy as large as $40$ MeV.

Van Beveren and Rupp \cite{rupp} argued from the experience with
$a_0/f_0 (980)$ that the low mass of $D_{sJ}(2317)$ could arise
from the mixing between $D K$ continuum and the lowest scalar
nonet. In this way the $0^+$ $\bar c s$ state is artificially
pushed much lower than that expected from quark models.

The recent lattice calculation suggests a value around $2.57$ GeV
for $0^+$ state mass \cite{bali}, much larger than the
experimentally observed $D_{sJ}(2317)$ and compatible with quark
model predictions. The conclusion in Ref. \cite{bali} is that
$D_{sJ}(2317)$ might receive a large component of $D K$ and the
physics might resemble $a_0/f_0(980)$. Such a large $D K$
component makes lattice simulation very difficult.

In this paper we shall use QCD sum rules \cite{svz} in the
framework of the heavy quark effective theory (HQET)
\cite{grinstein} to extract the masses since HQET provides a
systematic method to compute the properties of heavy hadrons
containing a single heavy quark via the $1/m_Q$ expansion, where
$m_Q$ is the heavy quark mass.  The masses of ground state heavy
mesons have been studied with QCD sum rules in HQET in
\cite{bagen,ball,luo}.  In \cite{huang,daizhu,colangelo98} masses
of the lowest excited non-strange heavy meson doublets $(0^+,
1^+)$ and $(1^+, 2^+)$ were studied with the sum rules in HQET up
to the order of ${\cal O}(1/m_Q)$. These masses were also analyzed
in the earlier works \cite{add} with sum rules in full QCD . In
this short note we extend the same formalism in
\cite{huang,daizhu} to include the light quark mass in calculating
the $\bar c s$ mesons. We shall also use more stringent criteria
for the stability windows of the sum rules in the numerical
analysis.

\section{$m_M-m_Q$ at the leading order of HQET}
\label{sec2}

The proper interpolating current $J_{j,P,j_{\ell}}^{\alpha_1\cdots\alpha_j}$
for a state with the quantum numbers $j$, $P$, $j_{\ell}$ in HQET was
given in \cite{huang}. These currents were proved to satisfy the following
conditions
\begin{eqnarray}
\label{decay}
\langle 0|J_{j,P,j_{\ell}}^{\alpha_1\cdots\alpha_j}(0)|
j',P',j_{\ell}^{'}\rangle&=& f_{Pj_l}\delta_{jj'}
\delta_{PP'}\delta_{j_{\ell}j_{\ell}^{'}}\eta^{\alpha_1\cdots\alpha_j}\;,\\
\label{corr}
i\:\langle 0|T\left (J_{j,P,j_{\ell}}^{\alpha_1\cdots\alpha_j}(x)
J_{j',P',j_{\ell}'}^{\dag\beta_1\cdots\beta_{j'}}(0)\right )|0\rangle&=&
\delta_{jj'}\delta_{PP'}\delta_{j_{\ell}j_{\ell}'}(-1)^j\:{\cal S}\:
g_t^{\alpha_1\beta_1}\cdots g_t^{\alpha_j\beta_j}\nonumber\\[2mm]&&\times\:
\int \,dt\delta(x-vt)\:\Pi_{P,j_{\ell}}(x)
\end{eqnarray}
in the $m_Q\to\infty$ limit, where $\eta^{\alpha_1\cdots\alpha_j}$ is the
polarization tensor for the spin $j$ state, $v$ is the velocity of the heavy
quark, $g_t^{\alpha\beta}=g^{\alpha\beta}-v^{\alpha}v^{\beta}$ is the
transverse metric tensor, ${\cal S}$ denotes symmetrizing the indices and
subtracting the trace terms separately in the sets $(\alpha_1\cdots\alpha_j)$
and $(\beta_1\cdots\beta_{j})$, $f_{P,j_{\ell}}$ and $\Pi_{P,j_{\ell}}$ are
a constant and a function of $x$ respectively which depend only on $P$ and
$j_{\ell}$.

We consider the correlator
\begin{equation}\label{cor-1}
\Pi (\omega )=i\int d^4 x e^{ikx} \langle 0|T\left
(J_{j,P,j_{\ell}}^{\alpha_1\cdots\alpha_j}(x)J_{j',P',j_{\ell}'}^{\dag
\beta_1\cdots\beta_{j'}}(0)\right )|0\rangle\;
\end{equation}
where $\omega=2v\cdot k$.  It can be written as
\begin{equation}
\Pi (\omega )= {f^2_{P,j_{\ell}} \over {{2\bar
\Lambda}_{j,P,j_{\ell}}-\omega}} + \mbox{higher states} \;,
\end{equation}
Where
$\bar\Lambda_{j,P,j_{\ell}}=lim_{m_Q\rightarrow\infty}(m_{M_{j,P,j_{\ell}}}
-m_Q)$. On the other hand, it will be calculated in terms of
quarks and gluons. Invoking Borel transformation to Eq.
(\ref{cor-1}) we get
\begin{equation}
f^2_{P,j_{\ell}} e^{-{2\bar\Lambda_{j,P,j_{\ell}}\over
T}}={1\over \pi}\int_{2m_q}^{\omega_c} \rho (\omega)
e^{-{\omega\over T}} + \mbox{condensates} \;.
\end{equation}
where $m_q$ is the light quark mass, $\rho (\omega)$ is the
perturbative spectral density and $\omega_c$ is the threshold
parameter used to subtract the higher state contribution with the
help of quark-hadron duality assumption.

We shall study the low-lying $(0^+, 1^+), (1^+, 2^+)$ $(\bar c
s)$ states. The values of various QCD condensates are
\begin{eqnarray}
\label{parameter}
\langle\bar ss\rangle&=&-(0.8\pm 0.1)*(0.24 ~\mbox{GeV})^3\;,\nonumber\\
\langle\alpha_s GG\rangle&=&0.038 ~\mbox{GeV}^4\;,\nonumber\\
m_0^2&=&0.8 ~\mbox{GeV}^2\;.
\end{eqnarray}
We use $m_s (1 \mbox{GeV}) =0.15$ GeV for the strange quark mass
in the ${\bar {MS}}$ scheme. We use $\Lambda_{QCD}=375$ MeV for
three active flavors and $\Lambda_{QCD}=220$ MeV for four active
flavors. The sum rules with massless light quarks have been
obtained in \cite{daizhu,huang}.

In the numerical analysis of the QCD sum rules we require that the
high-order power corrections be less than $30\%$ of the
perturbation term.  This condition yields the minimum value
$T_{min}$ of the allowed Borel parameter. We also require that the
 pole term, which is equal to the sum of the
cut-off perturbative term and the condensation terms, is larger
than $60\%$ of the perturbative term, which leads to the maximum
value $T_{max}$ of the allowed $T$. Thus we have the working
interval $T_{min} < T < T_{max}$ for a fixed $\omega_c$. If
$T_{min}\ge T_{max}$, we are unable to extract useful information
from such a sum rule. In the ideal case, the difference between
the meson mass $m_M$ and the heavy quark mass $m_Q$ (or other
observables) is almost independent on $T$ for certain values of
$\omega_c$. Namely, the dependence on both the Borel parameter and
continuum threshold is minimum. In realistic cases, the variation
of a sum rule with both $T$ and $\omega_c$ will contribute to the
errors of the extracted value, together with the truncation of the
operator product expansion and the uncertainty of vacuum
condensate values.

In subsections 2.1 and 2.2 we give sum rules for $(0^+, 1^+)$ and
$(1^+, 2^+)$ doublets using interpolating currents with
derivatives, respectively. For comparison we discuss sum rules for
$(0^+, 1^+)$ using interpolating currents without derivatives in
subsection 2.3.

\subsection{ $(0^+, 1^+)$ Doublet With Derivative Currents}

We employ the following interpolating currents \cite{huang}
\begin{eqnarray}\nonumber
J^{'\dag}_{0,+,2}&=&\frac{1}{\sqrt{2}}\:\bar h_v(-i)\not\!{\cal D}_t s\;,\\
\label{curr4} J^{'\alpha\dag}_{1,+,2}&=&\frac{1}{\sqrt{2}}\:\bar
h_v\gamma^5\gamma^{\alpha}_t(-i)\not\!{\cal D}_t s\;.
\end{eqnarray}
where $D_t^\mu=D^\mu-(v\cdot D)\,v^\mu$, with
$D^\mu=\partial^\mu-i g\, A^\mu$ being the gauge-covariant
derivative.  We obtain the sum rule relevant to $\bar{\Lambda}$
\begin{eqnarray}
\label{form1} \nonumber
{f'}_{+,1/2}^2e^{-2\bar\Lambda/{T}}&=&
\frac{3}{64\pi^2}\int_{2m_s}^{\omega_c} [\omega^4+2m_s
\omega^3-6m_s^2 \omega^2-12 m_s^3 \omega] e^{-\omega/{T}}d\omega
\\
&&-\frac{1}{16}\,m_0^2\,\langle\bar ss\rangle+{3\over 8}m_s^2
\,\langle\bar ss\rangle-{m_s\over 16\pi}\,\langle \alpha_s G^2
\rangle\;.
\end{eqnarray}

\begin{figure}
\label{fig1}
\begin{center}
\epsfig{file=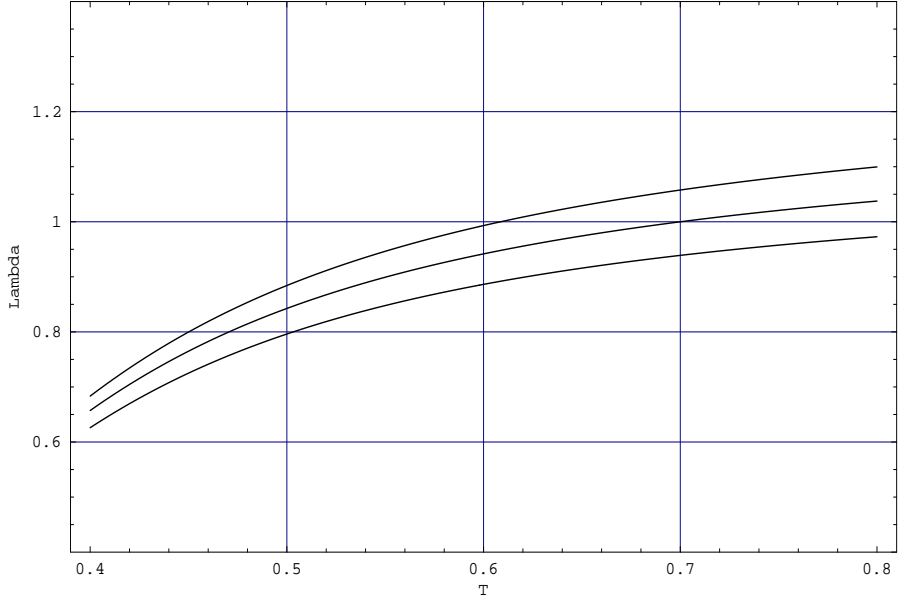,height=8cm,width=12cm} \caption{The variation
of $\bar\Lambda (\frac{1}{2}^+)$ (in unit of GeV) of the $(0^+,
1^+)$ doublet with $T$ and $\omega_c$ for the derivative currents.
The vertical and horizontal axes correspond to $\bar \Lambda$ and
$T$. From top to bottom, the curves correspond to $\omega_c$ being
$3.1, 2.9, 2.7$ GeV respectively.}
\end{center}
\end{figure}

Taking the derivative of the logarithm of the above equation with
respect to $(\frac{1}{T})$ one obtains the sum rule for
$\bar\Lambda$. Substituting the obtained value of $\bar\Lambda$ in
(\ref{form1}) one obtains the sum rule for ${f'}_{+,1/2}$. In
Figure 1, the variation of $\bar\Lambda({1/2}^+)$ with $T$ and
$\omega_c$ is shown. According to the criteria stated above the
working range is $0.38<T<0.58$ GeV. We have
\begin{eqnarray}\nonumber
\label{result1} \bar\Lambda({1/2}^+)=(0.86\pm 0.10) ~~\mbox{GeV},
&\\
\label{ff1} f'_{+, 1/2}=(0.31\pm 0.05) ~~\mbox{GeV}^{5/2},
\end{eqnarray}
where the central value corresponds to $T=0.52$ GeV and $\omega_c
=2.9$ GeV .

\subsection{ $(1^+, 2^+)$ Doublet }

For the $(1^+, 2^+)$ doublet, by using the following interpolating
currents \cite{huang}
\begin{eqnarray}
\label{curr5}
\nonumber
J^{\dag\alpha}_{1,+,1}&=&\sqrt{\frac{3}{4}}\:\bar
h_v\gamma^5(-i)\left(
{\cal D}_t^{\alpha}-\frac{1}{3}\gamma_t^{\alpha}\not\!{\cal D}_t\right)q\;,\\
\label{curr6}
J^{\dag\alpha_1,\alpha_2}_{2,+,1}&=&\sqrt{\frac{1}{2}}\:\bar h_v
\frac{(-i)}{2}\left(\gamma_t^{\alpha_1}{\cal
D}_t^{\alpha_2}+\gamma_t^{\alpha_2}{\cal
D}_t^{\alpha_1}-\frac{2}{3}\;g_t^{\alpha_1\alpha_2}\not\!{\cal
D}_t\right)q\;,
\end{eqnarray}
we obtain the sum rule
\begin{eqnarray}
\label{form3}\nonumber
 f_{+,3/2}^2e^{-2\bar\Lambda/{T}}&=&{1\over
64\pi^2}\int_{2m_s}^{\omega_c}[\omega^4+2m_s\omega_c^3-6m_s^2
\omega^2-12 m_s^3 \omega] e^{-\omega/{T}}d\omega\\
&&-\frac{1}{12}\:m_0^2\:\langle\bar ss\rangle-{1\over
32}\langle{\alpha_s\over\pi}G^2\rangle T +{1\over
8}m_s^2\langle\bar ss\rangle-{m_s\over
48\pi}\langle{\alpha_s\over\pi}G^2\rangle\;.
\end{eqnarray}
\begin{figure}
\label{fig2}
\begin{center}
\epsfig{file=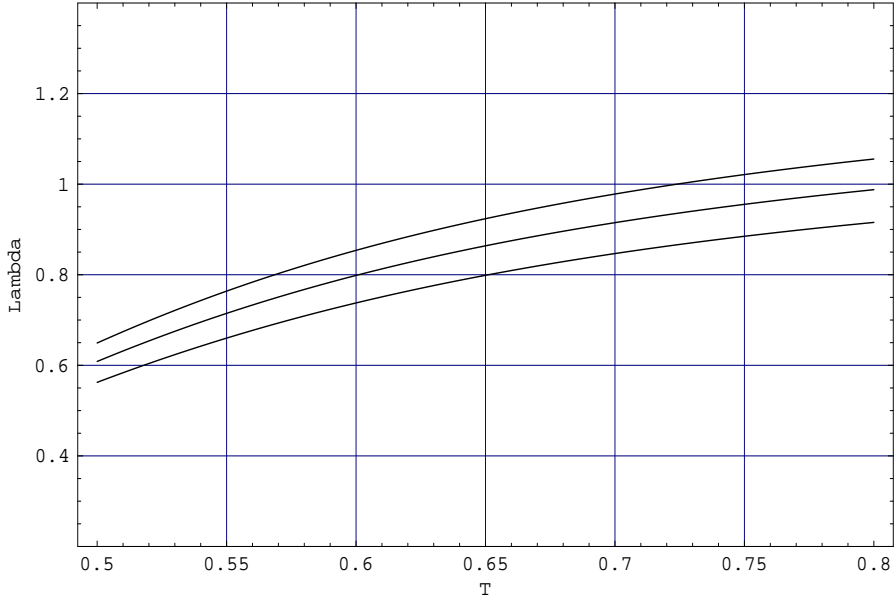,height=8cm,width=12cm} \caption{The variation
of $\bar \Lambda (\frac{3}{2}^+)$ (in unit of GeV) of the $(1^+,
2^+)$ doublet with $T$ and $\omega_c$. From top to bottom,
$\omega_c =3.2, 3.0, 2.8$ GeV respectively.}
\end{center}
\end{figure}

>From the sum rule, the variation of $\bar \Lambda({3/2}^+)$  with $T$
and $\omega_c$ is plotted in Figure 2, . The working range is
$0.55<T<0.65$ GeV.
We have
\begin{eqnarray}
\label{result2}\nonumber
 \bar\Lambda({3/2}^+)=(0.83\pm 0.10)
~~\mbox{GeV},
&\\
\label{ff2} f_{+, 3/2}=(0.19\pm 0.03) ~~\mbox{GeV}^{5/2},
\end{eqnarray}
where the central value corresponds to $T=0.62$ GeV and $\omega_c
=3.0$ GeV .

\subsection{ $(0^+, 1^+)$ Doublet With Currents Without Derivative}

Possible different currents for $(0^+, 1^+)$ doublet are the
non-derivative currents
\begin{eqnarray}
\label{curr1}\nonumber
J^{\dag}_{0,+,2}&=&\frac{1}{\sqrt{2}}\:\bar h_vs\;,\\
\label{curr2} J^{\dag\alpha}_{1,+,2}&=&\frac{1}{\sqrt{2}}\:\bar
h_v\gamma^5\gamma^{\alpha}_t s\;.
\end{eqnarray}

With the non-derivative currents the sum rule reads
\begin{eqnarray}
\label{form2}\nonumber
 f_{+,1/2}^2e^{-2\bar\Lambda/{T}}&=&
\frac{3}{16\pi^2}\int^{\omega_c}_{2m_s} [\omega^2-2 m_s \omega
-2m_s^2] e^{-\omega/{T}}d\omega \\&&+\frac{1}{2}\,\langle\bar
ss\rangle+\frac{m_s}{4T}\,\langle\bar ss\rangle -{1\over
8T^2}\,m_0^2\,\langle\bar ss\rangle\;.
\end{eqnarray}

Requiring that the condensate contribution is less than $30\%$ of
the perturbative term, we get $T_{\mbox{min}}=0.75$ GeV. In Figure
3, we show the ratio of the pole term, i.e the sum of the cut-off
perturbative term and condensation terms,
 for the sum rule (\ref{form2}). In the whole range of
$T>T_{\mbox{min}}$, this pole contribution is less than $40\%$.
Hence, There is no stability window for this sum rule satisfying
our criteria stated before.

\begin{figure}
\label{fig3}
\begin{center}
\epsfig{file=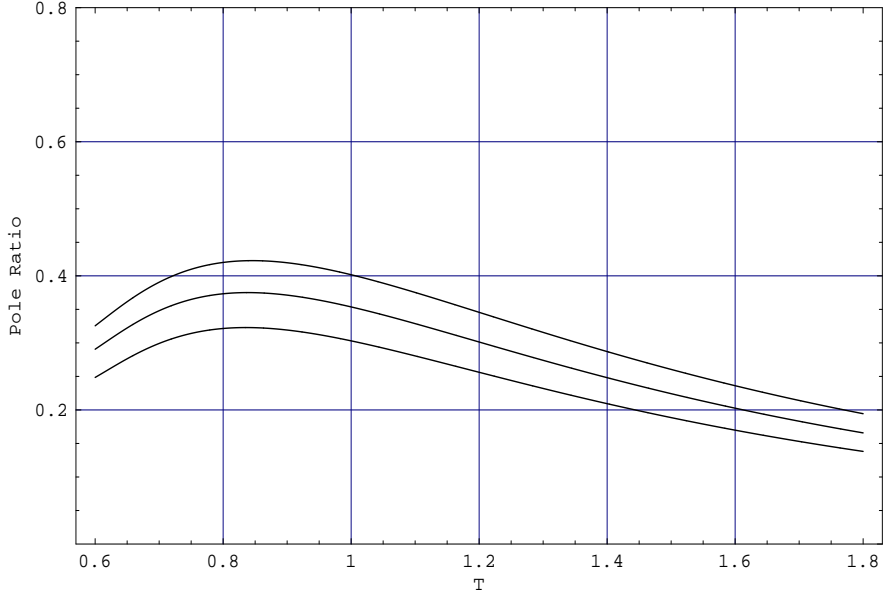,height=8cm,width=12cm} \caption{The variation
of the ratio of the pole contribution and the perturbative piece
of the $(0^+, 1^+)$ doublet with $T$ and $\omega_c$ for the
non-derivative currents. From top to bottom, $\omega_c =3.1, 2.9,
2.7$ GeV respectively.}
\end{center}
\end{figure}

If we arbitrarily loosen the analysis criteria and require the
pole contribution is greater than $30\%$ only, we get the working
range $0.75<T<1.2$ GeV. In Figure 4, the variation of
$\bar\Lambda({1/2}^+)$ with $T$ and $\omega_c$ is shown.
Numerically,
\begin{eqnarray}
\label{result3}\nonumber \bar\Lambda({1/2}^+)=(1.30\pm
0.15)~~\mbox{GeV},
&\\
\label{ff3} f_{+, 1/2}=(0.39\pm 0.05 )~~\mbox{GeV}^{3/2},
\end{eqnarray}
where the central value corresponds to $T=1.0$ GeV and $\omega_c
=2.9$ GeV. Because of the weaker criteria used these results are
less reliable and will not be used in the final numerical results.

Here we would like to make some remarks.  Usually, the currents
with the least number of derivatives are used in QCD sum rule
approaches.  The sum rules then are less sensitive to the
threshold energy $\omega_c$. However, it is pointed out in
\cite{huang} that in the non-relativistic limit the coupling
constant of these currents to the P wave states vanishes. If this
coupling constant is suppressed due to this reason, the relative
importance of the contribution of the $DK$ and other states in
continuum in the sum rules which are usually neglected would be
enhanced.  Besides, it is shown in \cite{blok,zhu1} by using soft
pion theorem that the contribution of the $D\pi$ continuum is
large in the sum rule with the non-derivative current for the
$0^+$ state of the non-strange $D$ system and significantly
decreases the value of $\bar\Lambda$. Similar method of
calculation is not good in the present case, but it indicates that
the contribution of the $DK$ continuum with the non-derivative
current may be large too.

\begin{figure}
\label{fig4}
\begin{center}
\epsfig{file=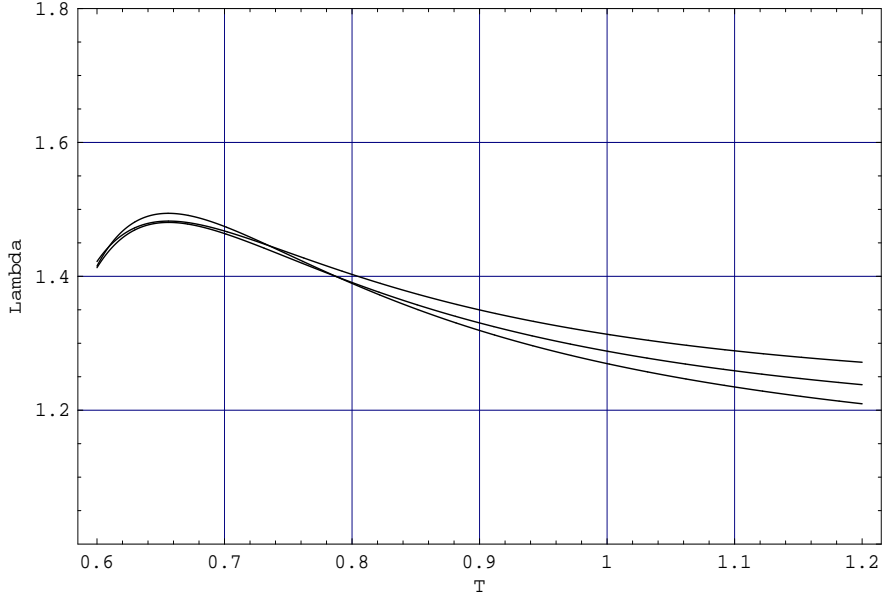,height=8cm,width=12cm} \caption{The variation
of $\bar \Lambda$ (in unit of GeV) of the $(0^+, 1^+)$ doublet
with $T$ and $\omega_c$ for the non-derivative currents. From top
to bottom, $\omega_c =3.1, 2.9, 2.7$ GeV respectively.}
\end{center}
\end{figure}

\section{Sum rules at the $1/m_Q$ order}
\label{sec3}

To the order of $1/m_Q$, the Lagrangian of HQET is
\begin{equation}
\label{Leff}
   {\cal L}_{\rm eff} = \bar h_v\,i v\!\cdot\!D\,h_v
   + \frac{1}{2 m_Q}\,{\cal K}
   + \frac{1}{2 m_Q}\,{\cal S}+{\cal O}(1/m_Q^2) \,,
\end{equation}
where $h_v(x)$ is the velocity-dependent effective field related to
the original heavy-quark field $Q(x)$ by
\begin{equation}
   h_v(x) = e^{i m_Q v\cdot x}\,\frac{1+\rlap/v}{2}\,Q(x)\;.
\end{equation}
 ${\cal K}$ is the kinetic operator defined as
\begin{equation}
\label{kinetic} {\cal K}=\bar h_v\,(i D_t)^2 h_v\;,
\end{equation}
$\cal S$ is the chromo-magnetic operator
\begin{equation}
\label{pauli} {\cal S}=\frac{g}{2}\,C_{mag}(m_Q/\mu)\;
   \bar h_v\,\sigma_{\mu\nu} G^{\mu\nu} h_v\;,
\end{equation}
where $C_{mag}=\displaystyle{\left(\alpha_s(m_Q)\over
\alpha_s(\mu)\right)^{3/{\beta}_0}}$, ${\beta}_0=11-2n_f/3$.

Considering ${\cal O}(1/m_Q)$ corrections
the pole term of the correlator on the hadron side becomes
\begin{eqnarray}
\label{m-pole}
\Pi(\omega)_{pole}={(f+\delta f)^2\over 2(\bar\Lambda+\delta m)-\omega}
                  ={f^2\over 2\bar\Lambda-\omega}
                  -{2\delta mf^2\over (2\bar\Lambda-\omega)^2}
                  +{2f\delta f\over 2\bar\Lambda-\omega}\;,
\end{eqnarray}
where $\delta m$ and $\delta f$ are of order ${\cal O}(1/m_Q)$.

To extract $\delta m$ in (\ref{m-pole})
we follow the approach of \cite{ball} to consider
the three-point correlation functions
\begin{eqnarray}
\label{delta}
\delta_O\Pi^{\alpha_1\cdots\alpha_j,\beta_1\cdots\beta_j}_{j,P,i}
(\omega,\omega')
=i^2\int d^4xd^4ye^{ik\cdot x-ik'\cdot y}\langle 0|T
\left(J_{j,P,i}^{\alpha_1\cdots\alpha_j}(x)\;O(0)\;
J_{j,P,i}^{\dag\beta_1\cdots\beta_{j}}(y)\right )|0\rangle\;,
\end{eqnarray}
where $O={\cal K}$ or ${\cal S}$. The scalar function
corresponding to (\ref{delta}) can be represented as the double
dispersion integral
\begin{eqnarray}\nonumber
\label{delta1}
\delta_O\Pi(\omega,\omega')
={1\over\pi^2}\int{{\rho}_o(s,s')dsds'\over(s-\omega)(s'-\omega')}\;.
\end{eqnarray}
The pole parts of $\delta_O\Pi(\omega,\omega'), O=\cal K, \cal S$
are
\begin{eqnarray}
\label{polek}
&&\delta_{\cal K}\Pi(\omega,\omega')_{pole}=
{f^2K\over(2\bar\Lambda-\omega)(2\bar\Lambda-\omega')}
+{f^2G_{\cal K}(\omega')\over 2\bar\Lambda-\omega}
+{f^2G_{\cal K}(\omega)\over 2\bar\Lambda-\omega'}\;,\\[2mm]
\label{poles}
&&\delta_{\cal S}\Pi(\omega,\omega')_{pole}=
{d_Mf^2\Sigma\over(2\bar\Lambda-\omega)(2\bar\Lambda-\omega')}
+d_Mf^2\left[{G_{\cal S}(\omega')\over 2\bar\Lambda-\omega}
+{G_{\cal S}(\omega)\over 2\bar\Lambda-\omega'}\right]\;,
\end{eqnarray}
where
\begin{eqnarray}\nonumber
&&{K_{j,P,j_l}}=\langle j,P,j_l|{\bar h_v\,(i D_\perp)^2
h_v}|j,P,j_l\rangle\;,\\ \nonumber
&&{2d_M\Sigma_{j,P,j_l}}=\langle j,P,j_l|{
   \bar h_v\,g\sigma_{\mu\nu} G^{\mu\nu} h_v}|j,P,j_l\rangle\;,\\
   \nonumber
&&d_M=d_{j,j_l},\hspace{2.5mm}d_{j_l-1/2,j_l}
=2j_l+2,\hspace{2.5mm}d_{j_l+1/2,j_l}=-2j_l.
\end{eqnarray}
Letting $\omega=\omega'$ in eqs. (\ref{polek}) and (\ref{poles})
and comparing with (\ref{m-pole}), one obtains \cite{ball}
\begin{eqnarray}
\label{delm}
\delta m=-{1\over 4m_Q}(K+d_MC_{mag}\Sigma)\;.
\end{eqnarray}
The single pole terms in (\ref{polek}) and (\ref{poles}) come from
the region in which $s (s')=2\bar\Lambda$ and $s' (s)$ is at the
pole for a radial excited state or in the continuum. They are
suppressed by making the double Borel transformation for both
$\omega$ and $\omega'$.  The Borel parameters corresponding to
$\omega$ and $\omega'$ are taken to be equal.  One obtains thus
the sum rules for $K$ and $\Sigma$ as
\begin{eqnarray}
\label{form4}
&&f^2K\,e^{-2\bar\Lambda/{T}}=\int_{2m_s}^{\omega_c}\int_{2m_s}^{\omega_c}
d\omega d\omega'e^{-(\omega+\omega')/2T}\rho_{\cal K}(\omega,\omega')\;,\\
&&f^2\Sigma\,e^{-2\bar\Lambda/{T}}
=\int_{2m_s}^{\omega_c}\int_{2m_s}^{\omega_c}
d\omega d\omega'e^{-(\omega+\omega')/2T}\rho_{\cal
S}(\omega,\omega')\;.
\end{eqnarray}

In this section we shall neglect the $m_s$ corrections to $K$ and
$\Sigma$. We obtain for the $j_l^P=\displaystyle{{1\over 2}^+}$
doublet with derivative currents
\begin{eqnarray}
\label{form-k1d}\nonumber
&&f_{+,1/2}^{'2}K\,e^{-2\bar\Lambda/{T}}= -{3\over
2^7\pi^2}\int_{2m_s}^{\omega_c} \omega^6\;
e^{-\omega/T}d\omega+{3\over 2^4\pi}\langle\alpha_sGG\rangle\;T^3\;,\\
\label{form-s1d} &&f_{+,1/2}^{'2}\Sigma\,e^{-2\bar\Lambda/{T}}=
\int_{2m_s}^{\omega_c}\int_{2m_s}^{\omega_c} \rho (s, s^\prime
)\;e^{-{s+s^\prime \over 2T}}dsds^\prime +{1\over
48\pi}\langle\alpha_sGG\rangle \;T^3\;,
\end{eqnarray}
with
\begin{eqnarray} \nonumber
\rho (s,s^\prime )={1\over 96}{\alpha_s(2T)\over \pi^3} C_{mag}
ss^\prime \{ {s^\prime}^2 (3s-s^\prime ) \theta (s-s^\prime ) +
(s\leftrightarrow s^\prime ) \} \; .
\end{eqnarray}

\begin{figure}
\label{fig5}
\begin{center}
\epsfig{file=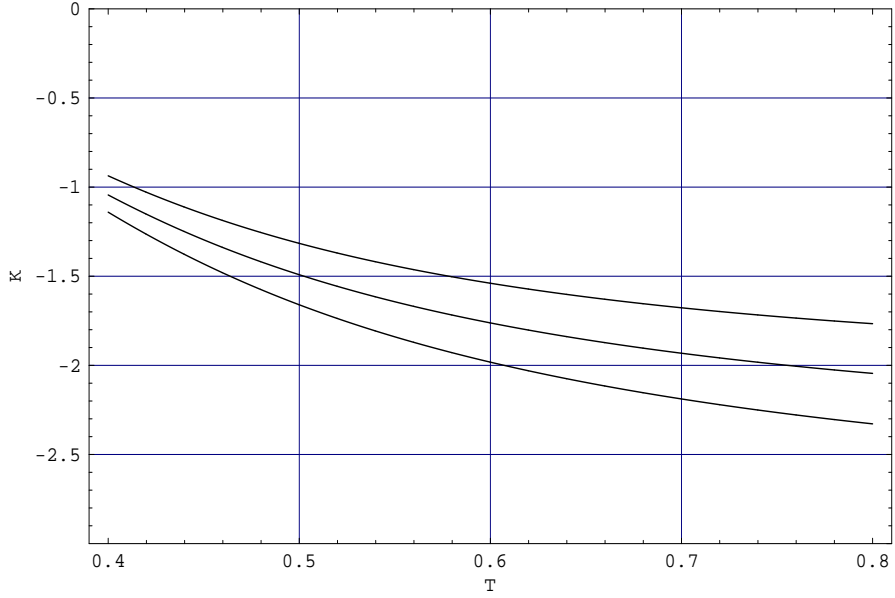,height=8cm,width=12cm} \caption{The variation
of $K_{1/2}$ (in unit of GeV$^2$) of the $(0^+, 1^+)$ doublet with
$T$ and $\omega_c$ for the derivative currents. From top to
bottom, $\omega_c =3.1, 2.9, 2.7$ GeV respectively.}
\end{center}
\end{figure}

We can eliminate the dependence  of the $K$ and $\Sigma$ on
$f_{+,1/2}$ and $\bar{\Lambda}$ through dividing the above sum
rules (\ref{form-k1d}) by the sum rule (\ref{form1}). We use the
same working windows as those of two-point sum rules for $\bar
\Lambda$ in evaluating ${\cal O}(1/m_Q)$ corrections. Numerically
we have
\begin{eqnarray}\label{k1}\nonumber
K_{1/2}=(-1.60\pm 0.30) ~~\mbox{GeV}^2\\
\Sigma_{1/2}=(0.28\pm 0.05) ~~\mbox{GeV}^2\;.
\end{eqnarray}
The variations of $K$ and $\Sigma$ with $T$ and $\omega_c$ for the
$j_l^P=\displaystyle{{1\over 2}^+}$ doublet are shown in Figure 5
and 6, respectively.

\begin{figure}
\label{fig6}
\begin{center}
\epsfig{file=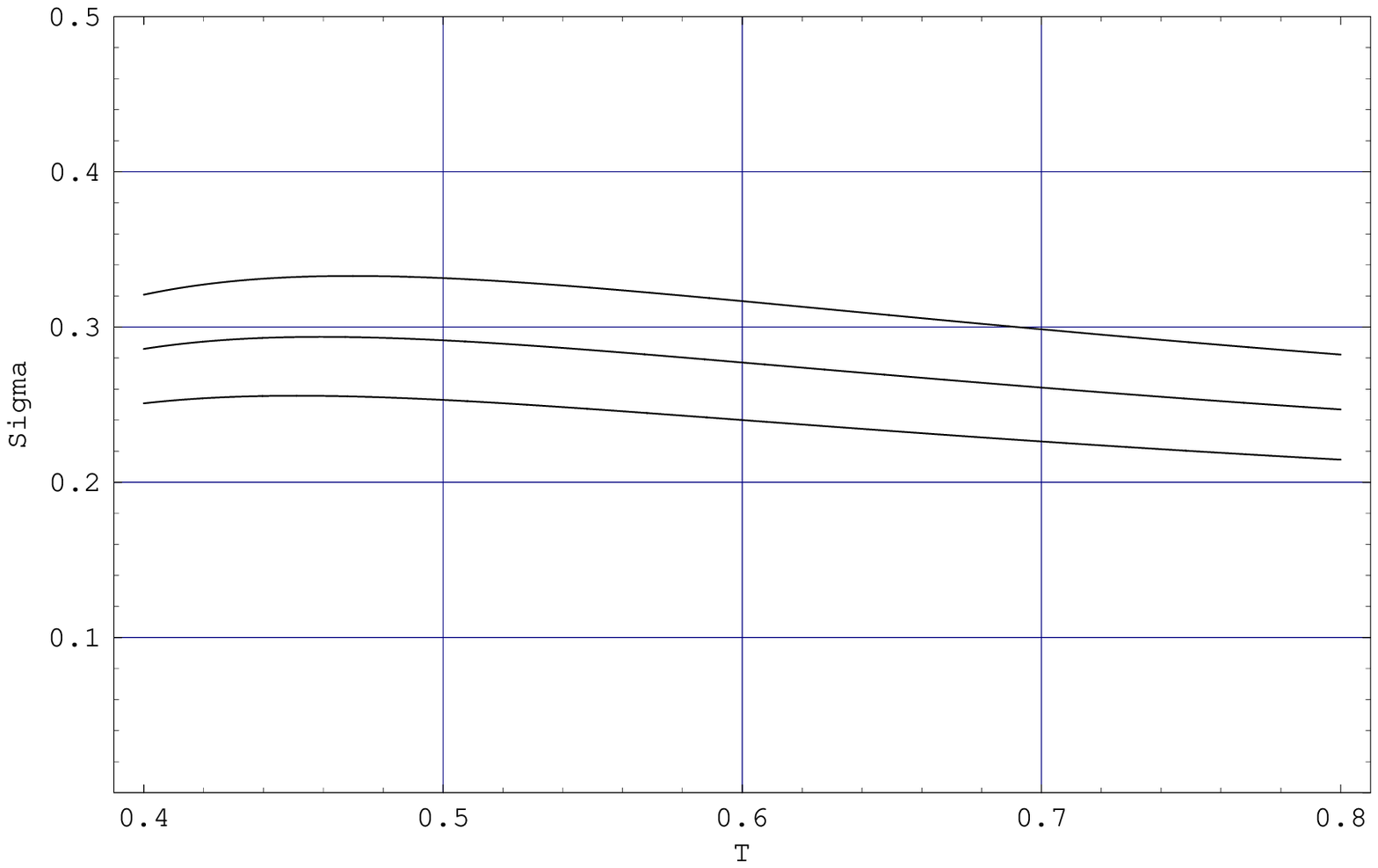,height=8cm,width=12cm} \caption{The variation
of $\Sigma_{1/2}$ (in unit of GeV$^2$) of the $(0^+, 1^+)$ doublet
with $T$ and $\omega_c$ for the derivative currents. From top to
bottom, $\omega_c =3.1, 2.9, 2.7$ GeV respectively.}
\end{center}
\end{figure}

For $j_l^P=\displaystyle{{3\over 2}^+}$ doublet, we have
\begin{eqnarray}\nonumber
\label{form-k2}
&&f_{+,3/2}^2K\,e^{-2\bar\Lambda/{T}}=-{1\over 2^7\pi^2}
\int_{2m_s}^{\omega_c} \omega^6\;e^{-\omega/T}
d\omega+{7\over 3\times 2^5\pi}\langle\alpha_sGG\rangle \;T^3\;,\\
\label{form-s2} &&f_{+,3/2}^2\Sigma\,e^{-2\bar\Lambda/{T}}=
\int_{2m_s}^{\omega_c}\int_{2m_s}^{\omega_c} \rho (s, s^\prime
)\;e^{-{s+s^\prime \over 2T}}dsds^\prime +{1\over
72\pi}\langle\alpha_sGG\rangle \;T^3\;.
\end{eqnarray}
with
\begin{eqnarray}\nonumber
\rho (s,s^\prime )={1\over 288}{\alpha_s(2T)\over \pi^3} C_{mag}
 \{ {s^\prime}^4 (s- {3\over 5}s^\prime ) \theta (s-s^\prime )
+ (s\leftrightarrow s^\prime ) \}\;.
\end{eqnarray}
The variations of $K$ and $\Sigma$ with $T$ and $\omega_c$ for the
$j_l^P=\displaystyle{{3\over 2}^+}$ doublet are plotted in Figure
7 and 8 respectively. Numerically we have
\begin{eqnarray}\nonumber\label{k2}
K_{3/2}=(-1.64\pm 0.40) ~~\mbox{GeV}^2\\
\Sigma_{3/2}=(0.058\pm 0.01) ~~\mbox{GeV}^2\;.
\end{eqnarray}

\begin{figure}
\label{fig7}
\begin{center}
\epsfig{file=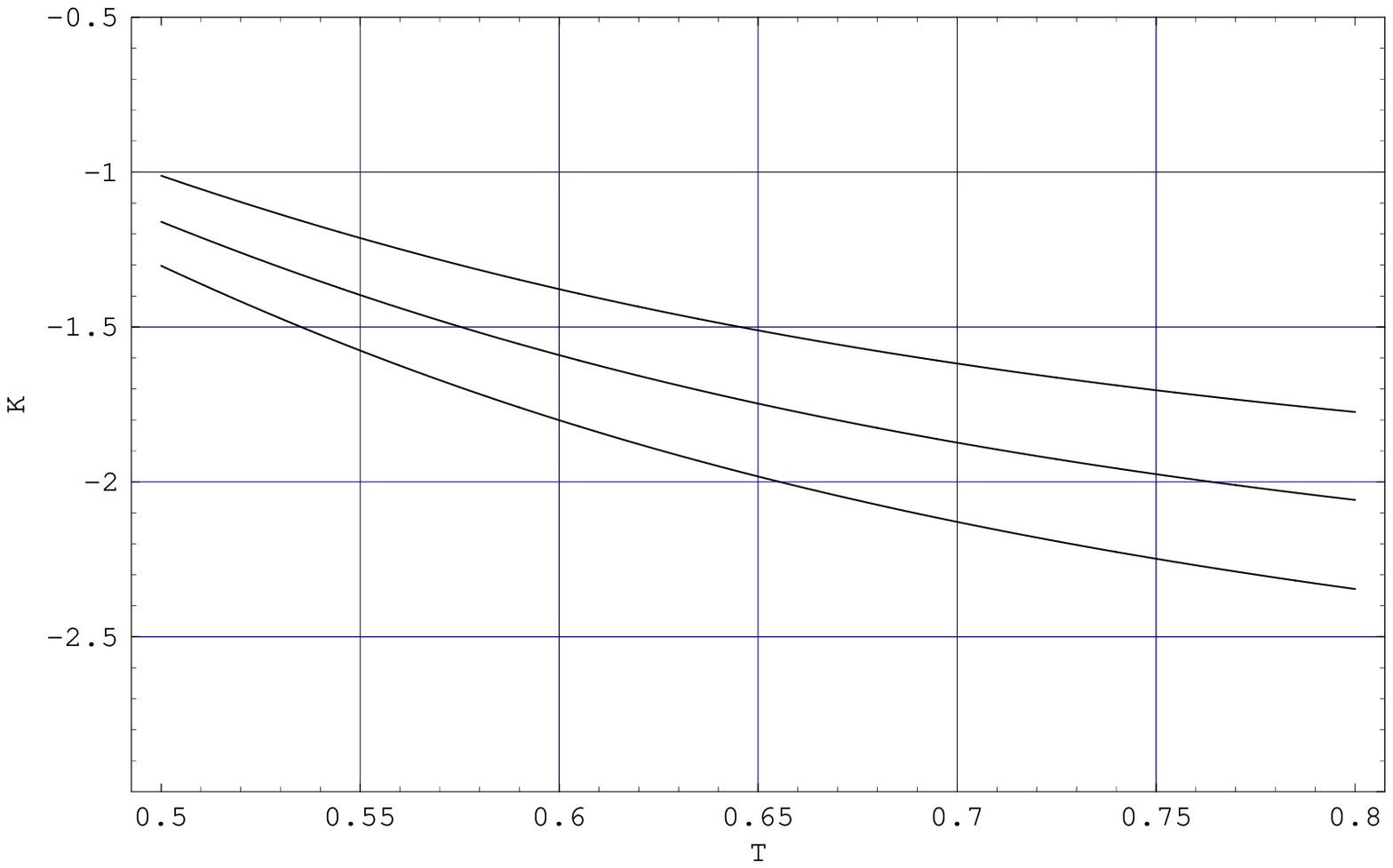,height=8cm,width=12cm} \caption{The variation
of $K_{3/2}$ (in unit of GeV$^2$) of the $(1^+, 2^+)$ doublet with
$T$ and $\omega_c$. From top to bottom, $\omega_c =3.2, 3.0, 2.8$
GeV respectively.}
\end{center}
\end{figure}

\begin{figure}
\label{fig8}
\begin{center}
\epsfig{file=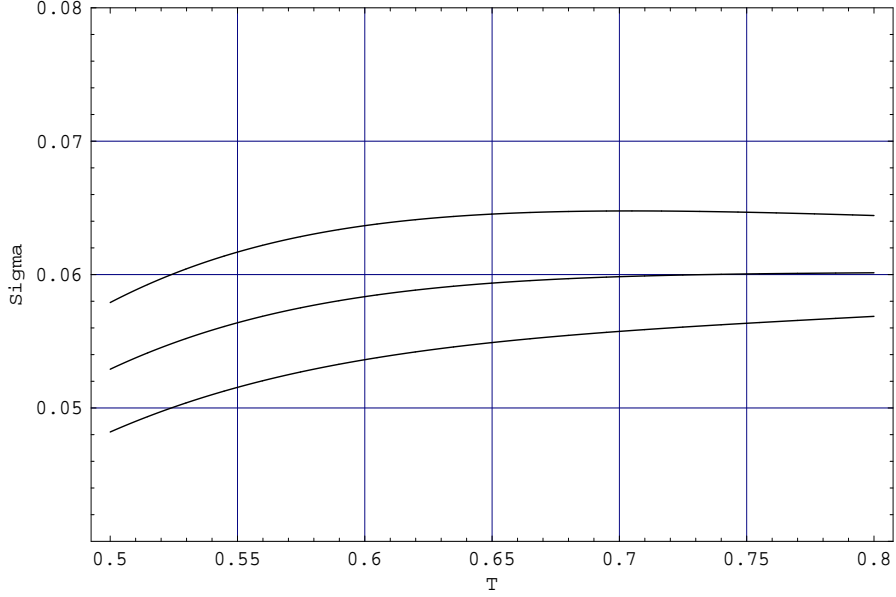,height=8cm,width=12cm} \caption{The variation
of $\Sigma_{3/2}$ (in unit of GeV$^2$) of the $(1^+, 2^+)$ doublet
with $T$ and $\omega_c$. From top to bottom, $\omega_c =3.2, 3.0,
2.8$ GeV respectively.}
\end{center}
\end{figure}

The spin-symmetry violating term ${\cal S}$ not only causes
splitting of masses within the same doublet, but also causes
mixing of states with the same $j$, $P$ but different $j_l$. In
Ref. \cite{daizhu} corrections from the mixing is found to be
negligible. We omit this effect here.

\section{Numerical Results and Discussions}
\label{sec5}

We present our results for the $\bar c s$ system assuming the HQET
is good enough for excited $D$ mesons. For the doublet $(0^+,1^+)$
with the derivative current, the weighted average mass is
\begin{eqnarray}\nonumber
 {1\over 4}\;(m_{D_{s0}^\ast}+3m_{D^*_{s1}})
=m_c+(0.86\pm 0.10)+{1\over m_c}\;[(0.40\pm 0.08)~\mbox{GeV}^2] .
\end{eqnarray}
The mass splitting is
\begin{eqnarray}\nonumber
m_{D^*_{s1}}-m_{D_{s0}^\ast}&=&{1\over m_c}\;[(0.28\pm
0.05)~\mbox{GeV}^2]\;.
\end{eqnarray}

For the $(1^+, 2^+)$ doublet we have
\begin{eqnarray}\nonumber
{1\over 8}\;(3m_{D_{s1}}+5m_{D_{s2}^*})&=&
m_c+(0.83\pm 0.10)+{1\over m_c}\;[(0.41\pm 0.10)~\mbox{GeV}^2]\;.\\
\end{eqnarray}
The $1^+, 2^+$ mass splitting is
\begin{eqnarray}\nonumber
m_{D_{s2}^*}-m_{D_{s1}}={1\over m_c}\;[(0.116\pm
0.06)~\mbox{GeV}^2]\;.
\end{eqnarray}

The results for $\bar b s$ system are obtained by replacing $m_c$
by $m_b$ and multiplying $\Sigma$ by $0.8$ (since $C_{mag} \approx
0.8$ for B system by using the values of $\Lambda_{\rm QCD}$
given in Sect. 2) in above equations.

Choosing $m_c$ to fit the experimental value
\begin{eqnarray}\nonumber
 {1\over 8}\;(3m_{D_{s1}}+5m_{D_{s2}^*})=2.56~\mbox{GeV} ,
\end{eqnarray}
where we again neglect the mixing between two $1^+$ states, we
obtain $m_c=1.44~\mbox{GeV}$. Using this $m_c$ value we obtain the
following numerical results. The $1^+, 2^+$ mass splitting is
\begin{eqnarray}\nonumber
m_{D_{s2}^*}-m_{D_{s1}}&=&(0.080\pm 0.042)~\mbox{GeV}\;
\end{eqnarray}
which is consistent with the experimental value $37$ MeV within
the large theoretical uncertainty. Experimentally, the mass
splitting in the $(1^+, 2^+)$ doublet in $D_s$ system is almost
equal to that in $D$ system. This justifies neglecting the $m_s$
correction to the $\Sigma$ term in our calculation. For
$(0^+,1^+)$ doublet, the weighted average mass is
\begin{eqnarray}\nonumber
 {1\over 4}\;(m_{D_{s0}^\ast}+3m_{D^*_{s1}})
=(2.57\pm 0.12)~\mbox{GeV}.
\end{eqnarray}
The mass splitting is
\begin{eqnarray}
\label{split} m_{D^*_{s1}}-m_{D_{s0}^\ast}&=&(0.19\pm
0.04)~\mbox{GeV}\;.
\end{eqnarray}
Therefore, the $0^+$ mass is predicted to be $m_{0^+}=(2.42\pm
0.13)~\mbox{GeV}$. This is consistent with the experimental value
$2.317~\mbox{GeV}$, though the central value is $100~\mbox{MeV}$
larger than data. If the $1^+$ $D_s$ state of mass
$2.460~\mbox{GeV}$ found by CLEO~\cite{cleo,babar} is assigned as
the other member of the $(0^+,1^+)$ doublet, then the observed
mass splitting $0.143 ~\mbox{GeV}$ is consistent with
(\ref{split}). Notice that the $\Sigma_{1/2}$ value in Fig. 6
changes very slowly between $T=0.35-0.8$ GeV. Therefore the result
(\ref{split}) is insensitive to the stability window used.

We would like to note that the stability of the sum rules (in
particular, those for the K and $\Sigma$) obtained by us is not as
good as those for the ground states, as can be seen from the
figures. And for the charm flavor hadrons, the $1/m_c$ corrections
are significant. So the predictions on masses have a relatively
large uncertainties, which are estimated by given errors.

If we replace the strange quark condensate by up/down quark
condensate and $m_s$ by the zero up/down quark mass, we can
extract the non-strange  excited $D$ meson masses. For $(1^+,
2^+)$ doublet, the sum rule window is $0.57 < T < 0.67$ GeV for
$\omega_c =2.8-3.2$ GeV. The results are
\begin{eqnarray}\label{811} \nonumber
\Lambda_{3/2}=(0.73\pm 0.08) \mbox{GeV}\, , \\ \nonumber
K_{3/2}=-(1.6\pm 0.4)\mbox{ GeV}^2\, ,\\
\Sigma_{3/2} =(0.057\pm 0.01) \mbox{GeV}^2
\end{eqnarray}
with the central value at $T=0.62$ GeV. For $(0^+, 1^+)$ doublet,
the sum rule window is $0.46 < T < 0.6$ GeV for $\omega_c
=2.7-3.1$ GeV. The results are
\begin{eqnarray}\label{911} \nonumber
\Lambda_{1/2}=(0.79\pm 0.08) \mbox{GeV}\, , \\ \nonumber
K_{1/2}=-(1.57\pm 0.4)\mbox{ GeV}^2\, ,\\
\Sigma_{1/2} =(0.286\pm 0.05) \mbox{GeV}^2
\end{eqnarray}
with the central value at $T=0.53$ GeV. The effect of strange
quark is to make the mass of $D_s$ mesons a little bit larger, as
expected. Experimentally, Belle collaboration found
$m_{D_0}=(2290\pm 22\pm 20)$ MeV and $m_{D_1^\ast}=(2400\pm 30\pm
20)$ MeV recently \cite{belle} while CLEO collaboration found
$m_{D_1^\ast}=2400^{+48}_{-42}$ MeV \cite{cleo1}. The $j_{3/2}$
non-strange doublet mass is known precisely \cite{pdg},
$m_{D_2}=2460$ MeV and $m_{D_1}=2420$ MeV. Our results
(\ref{811})-(\ref{911}) are consistent with the experimental data
within theoretical uncertainty.

In summary, we have calculated the masses of the excited
$(0^+,1^+)$ and $(1^+,2^+)$ doublets for the $\bar c s$ system to
the $1/m_Q$ order in the HQET sum rules. The numerical results
imply that the $D_{sJ}(2317)$ and $D_{sJ}(2460)$ observed by BABAR
and CLEO can be consistently identified as the $(0^+,1^+)$ doublet
with $j_l={1\over 2}^+$ within the theoretical uncertainty.
Especially, the mass splitting in the $(0^+,1^+)$ doublet is
reproduced quite well. The repulsion between these states and the
$DK$ and $DK^*$ continuum may help to lower their masses. In the
framework of the sum rule this effect should come from the
contribution from the $DK$ and $DK^*$ continuum to the dispersion
integral.

\vspace{0.8cm} {\it Acknowledgments:\/} This project was supported
by the National Natural Science Foundation of China, BEPC Opening
Project, Ministry of Education of China, FANEDD and SRF for ROCS,
SEM.

\end{document}